\documentclass[pre,aps,floats,superscriptaddress,floatfix,preprint]{revtex4}
\usepackage{amssymb,amsmath}
\UseRawInputEncoding
\usepackage{amsmath,amssymb}
\usepackage{graphicx}
\usepackage{psfrag}
\usepackage{color}
\usepackage{soul}
\usepackage{dcolumn}
\usepackage{bm}
\usepackage[normalem]{ulem}

\def\beq{\begin{equation}}
\def\eeq{\end{equation}}
\def\bea{\begin{eqnarray}}
\def\eea{\end{eqnarray}}
\begin{document}
\title{

Reduced fluctuations: Surprising effects of noise cross correlations in a coupled, driven model\\
}
\author{Sudip Mukherjee}\email{sudip.bat@gmail.com}
\affiliation{Barasat Government College,
10, KNC Road, Gupta Colony, Barasat, Kolkata 700124,
West Bengal, India}

\begin{abstract}
We elucidate how the strong coupling  phases of a coupled driven model, originally proposed in S. Mukherjee, Phys. Rev. E {\bf 108}, 024219 (2023),
are affected by noise cross correlations in general dimensions $d$. This model has two dynamical variables, where one of the
variables is autonomous being independent of the other, whereas the second one depends explicitly
on the former. 
By employing model coupling theories, we study the strong coupling phase of model. We show that the scaling laws in the strong coupling phase of the second field depend strongly on the strength of the noise cross correlations: the roughness exponent of the second field varies continuously with the noise cross correlation amplitude. As the latter amplitude rises, the roughness exponent gradually decreases, suggestion a novel suppression of the fluctuations of the second field in the strong coupling phase by noise cross correlations. We discuss the phenomenological implications of our results.

\end{abstract}

\maketitle

Numerous natural driven systems are described physically using coupled dynamics of multiple degrees of freedom; forced magnetohydrodynamics (MHD)~\cite{mhd} and symmetric mixtures of a miscible binary fluid~\cite{bin-fluid} are notable examples. Turbulence in these systems has been studied using stochastically driven MHD~\cite{mhd-turbu} binary fluid equations of the velocity and concentration gradient~\cite{bin-fluid-turbu1,bin-fluid-turbu2}. In equilibrium systems, conditions of thermal equilibrium, such as in the form of a Fluctuation-Dissipation-Theorem (FDT)~\cite{chaikin}, guarantee that noise statistics have no bearing on the system's thermodynamic properties. For example, relaxational dynamics with and without a conservation law for the order parameter~\cite{halpin} refer to the same equal-time, thermal equilibrium properties. In contrast, a change in the noise statistics can lead to significantly different nonequilibrium steady states (NESS) in driven systems, as opposed to equilibrium systems in the absence of any FDT.
For example, there are significant differences in the universal features of the generalised conserved Kardar-Parisi-Zhang (CKPZ+)~\cite{ckpz+} and generalised molecular beam epitaxy (MBE+)~\cite{mbe+} equations, which differ only in their noises. 

One of the fundamental questions in nonequilibrium statistical mechanics is whether or not the introduction of noise cross-correlations in a stochastically driven coupled model can result in a new NESS. The scaling characteristics of the NESS are really impacted in certain cases by non-zero cross-correlations of the two noises in the two dynamical equations. No symmetry argument or physical theory can rule out the existence of such noise cross-correlations in driven systems.
The significance of noise cross-correlations has been investigated further using simpler reduction models. For example, in a nonconserved relaxational model for the complex scalar field, noise cross-correlations prove to be generally a crucial influence on the scaling properties of the model near its critical point~\cite{niladri-o2}. It has been demonstrated in Refs.~\cite{abbmhd,abfrey1,abfrey2} that noise cross-correlations can result in continually changing scaling exponents in the NESS using a coupled Burgers model first put out in~\cite{abfrey1}. Very recently, the effects of noise cross-correlations on the steady states of driven, nonequilibrium
systems, which are described by two stochastically driven dynamical variables, one autonomous and another second field dynamically coupled to the first one, in one dimension (1D) has been studied~\cite{sm-noise}. This study reveals the striking striking effects of noise crosscorrelations: Such cross-correlations can cause instabilities in models that would otherwise be stable in the absence of any cross-correlations, depending on the specifics of the nonlinear coupling between the dynamical fields. In particular, for sufficiently weak noise crosscorrelation amplitudes, Ref.~\cite{sm-noise} shows that noise crosscorrelations are {\em irrelevant} in the renormalisation group (RG) language, making the model statistically identical to the one without noise crosscorrelations. However, as soon as the noise crosscorrelation amplitudes exceed a finite threshold, the RG flows become unstable, signaling the existence of a perturbatively inaccessible strong coupling phase, reminiscent of the roughening transition found in
the Kardar-Parisi-Zhang (KPZ) equation~\cite{stanley} in dimensions greater than two. It is also speculated that noise crosscorrelations, when relevant, suppress the fluctuations of the second dynamical field in 1D, suggesting a counter intuitive intriguing possibility of {\em reduced fluctuations} in the strong coupling phases.

In the present work,  we revisit the issue of the effects of noise
crosscorrelations on the NESS of the coupled model proposed and studied in Ref.~\cite{sm-noise}. We focus on the strong coupling phases in general $d$-dimensions. We use mode self-consistent coupling theories (SCMC) to explore the strong coupling phases. We calculate the scaling exponents in $d$-dimensions within a one-loop SCMC scheme. We find that the roughness exponent of the second field  decreases monotonically as the strength  of the noise crosscorrelations rise, confirming the speculations made in Ref.~\cite{sm-noise}.

We now set up our model and use it to derive the above results. Any multi-variable system with stochastically driven dynamical equations for the dynamical variables can have noise crosscorrelations. We employ a specially designed $d$-dimensional model that works well for our objectives because we are attempting to address concerns of principles. It is a
straightforward generalisation of the 1D model studied in Ref.~\cite{sm-noise} and consists of two vector-valued dynamical fields ${\bf v}({\bf x}, t)$ and ${\bf b}({\bf x}, t)$. Similar to Ref.~\cite{sm-noise}, we consider $\bf v$ to be autonomous, meaning it is independent of  $\bf b$, and its dynamics is given by the well-known  Burgers equation~\cite{fns}. It is given by
\begin{equation}
 \frac{\partial {\bf v}}{\partial t}=\nu \nabla^2 {\bf v} + \frac{\lambda}{2}{\boldsymbol \nabla}v^2 + {\bf f},\label{burg-v}
\end{equation}
where $\nu>0$ is a diffusivity, $\lambda$ is a nonlinear coupling constant of arbitrary sign and ${\bf f}$ is a zero-mean, Gaussian-distributed stochastic force. The second field $\bf b$ satisfies an advection-diffusion equation
\begin{equation}
 \frac{\partial {\bf b}}{\partial t} = \mu \nabla^2 {\bf b} + \lambda_2 {\boldsymbol\nabla} ({\bf v\cdot b}) + {\bf g},\label{burg-b}
\end{equation}
where $\mu>0$ is a diffusivity, $\lambda_1$ is a nonlinear coupling constant of arbitrary sign and ${\bf g}$ is a zero-mean, Gaussian-distributed stochastic force. Equations~(\ref{burg-v}) and (\ref{burg-b}) are invariant under the Galilean transformation if $\lambda=\lambda_1$, which is our focus below. 

For our purposes, it is convenient to re-express Eqs.~(\ref{burg-v}) and (\ref{burg-b}) in terms of nonconserved ``height fields'' $h({\bf x},t)$ and $\phi({\bf x},t)$ through ${\bf v}\equiv {\boldsymbol \nabla} h$ and ${\bf b}\equiv {\boldsymbol \nabla} h$. Fields $h$ and $\phi$ obey
\begin{eqnarray}
 \frac{\partial h}{\partial t} &=& \nu \nabla^2 h + \frac{\lambda}{2} ({\boldsymbol\nabla} h)^2 + f_h,\label{kpz}\\
 \frac{\partial \phi}{\partial t} &=& \mu\nabla^2 \phi + \lambda_1({\boldsymbol\nabla}h)\cdot ({\boldsymbol\nabla}\phi) + f_\phi,\label{phieq}
\end{eqnarray}
where $f_h$ and $f_\phi$ are zero-mean, Gaussian white noises, which are related to ${\bf f}_v$ and ${\bf f}_b$ via
${\bf f}_v = {\boldsymbol\nabla}f_h$ and ${\bf f}_b= {\boldsymbol\nabla}f_\phi$, respectively. Equation~(\ref{kpz}) is nothing but the well-known KPZ equation for surface growth~\cite{stanley}, Eq.~(\ref{phieq}) has been considered in Ref.~\cite{ertas}. We now give the variances of the noises $f_h$ and $f_\phi$ in the Fourier space:
\begin{eqnarray}
 \langle f_h({\bf k},t)f_h({\bf k}',0)\rangle &=& 2D_h \delta(t),\\
 \langle f_\phi({\bf k},t)f_\phi({\bf k}',0)\rangle &=& 2D_\phi \delta(t),\\
 \langle f_h({\bf k},t)f_\phi({\bf k}',0)\rangle &=& 2D_\times ({\bf k}) \delta(t),\label{noise-corr}
\end{eqnarray}
with $D_\times (-{\bf k})=-D_\times ({\bf k})$ and $D_\times ({\bf k})^2 = D_\times^2$, a constant. The form of the noise crosscorrelations is dictated by the parity properties of ${\bf v}$ and ${\bf b}$; see Refs.~\cite{abbmhd,abfrey1,sm-noise}. Only nonzero noise crosscorrelations can lead  to non-zero crosscorrelation function of $h$ and $\phi$.

\begin{figure}[htb]
 \includegraphics[width=4cm]{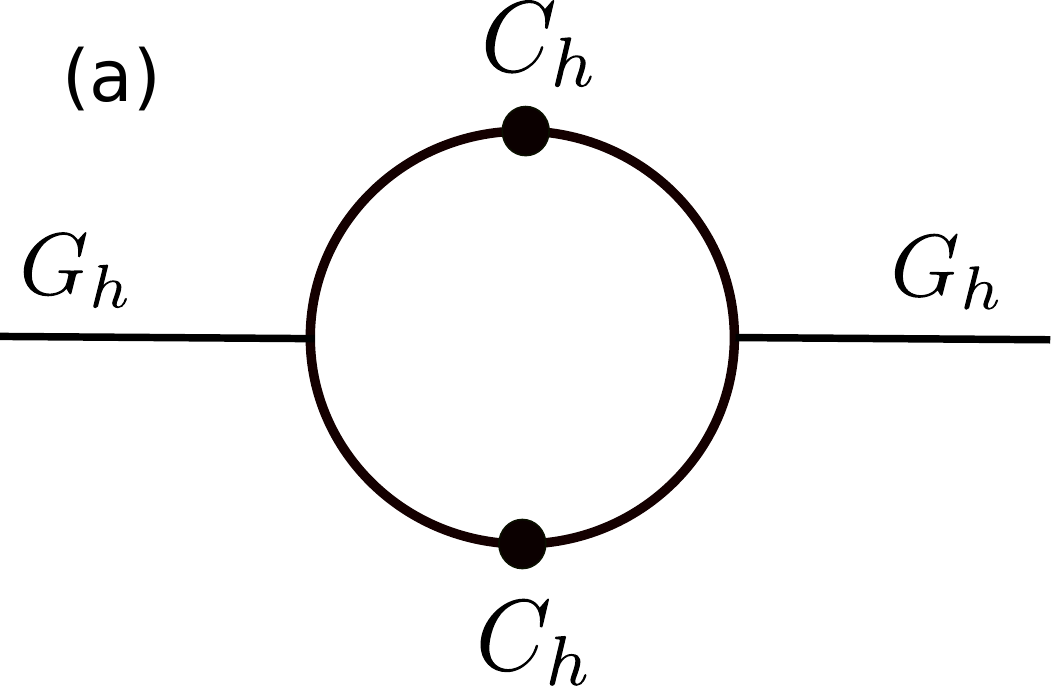}\hfill \includegraphics[width=4cm]{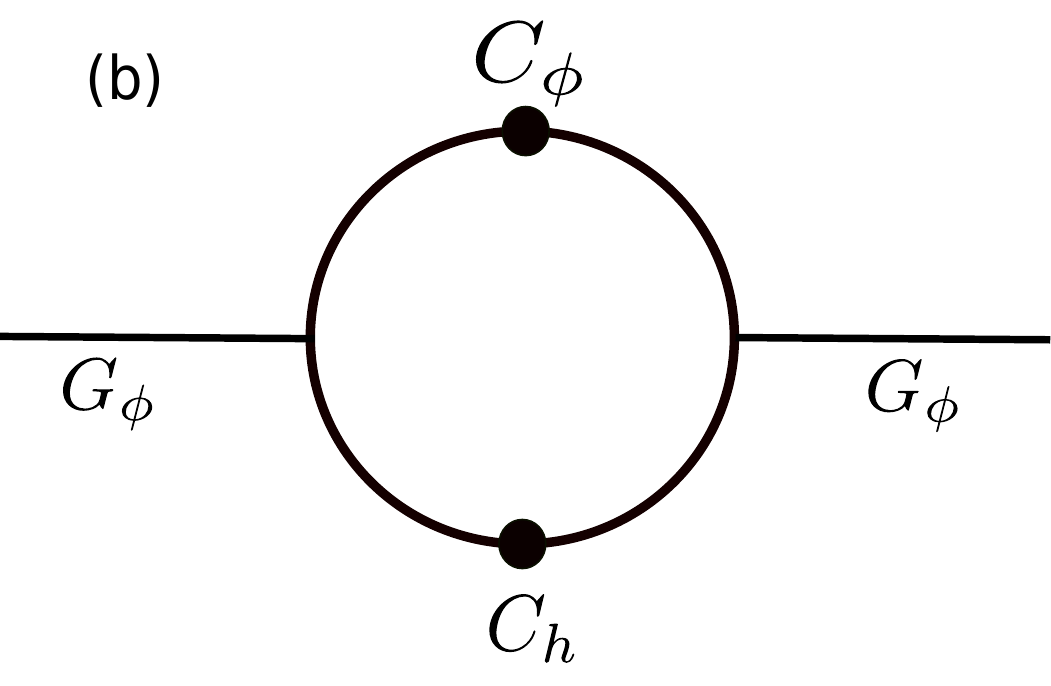}\\
 \includegraphics[width=4cm]{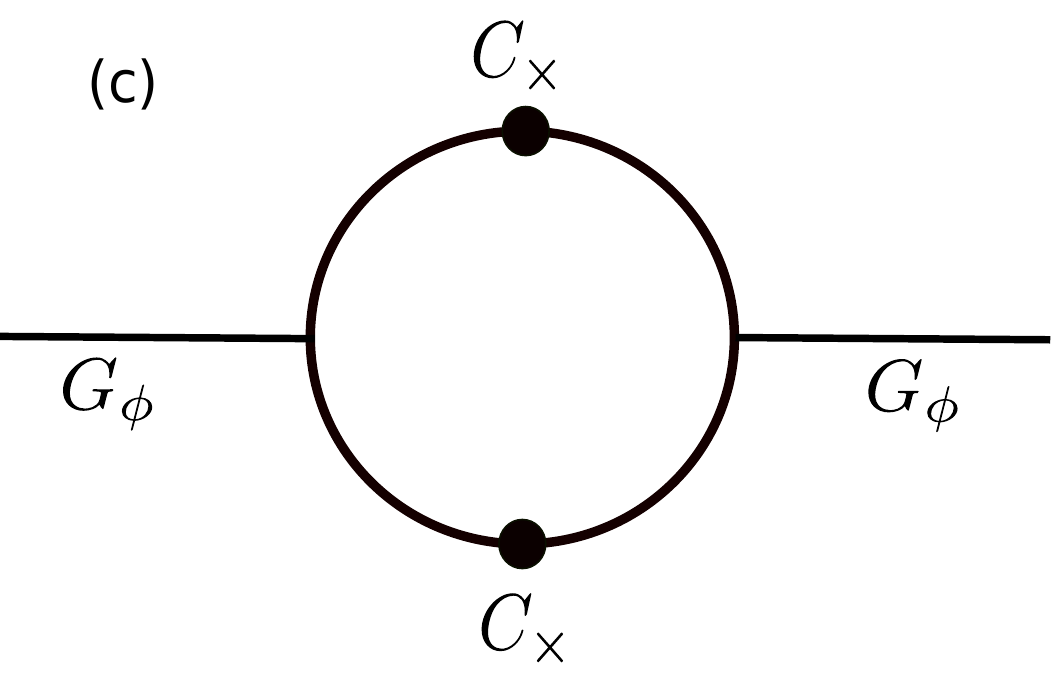}\hfill\includegraphics[width=2cm]{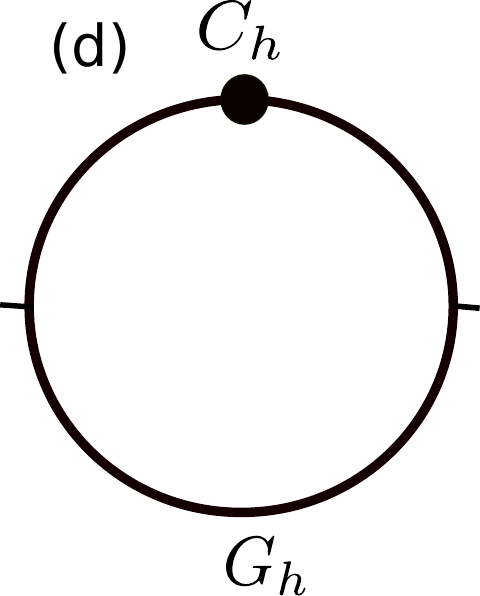} \hfill
 \caption{One-loop diagrams for (a) $C_h({\bf k},\omega)$, (b) $C_\phi({\bf k},\omega)$ that survives in the vanishing cross correlation limit, (c) $C_\phi({\bf k},\omega)$ that depends upon the cross correlations, (d) $\Gamma ({\bf k},\omega)$.}\label{diagrams}
\end{figure}

The KPZ equation (\ref{kpz}) in 2D has only a perturbatively inaccessible rough phase, whereas in $d>2$ there is a roughening transition at a finite threshold for the noise strength between a smooth and a perturbatively inaccessible rough phase. Failure of perturbative RG has led to development of alternative approaches to study the scaling properties in the rough phase. A notable among them is SCMC~\cite{tim,cates,jkb-mct,frey-mct}, which have predicted a variety of results, in spite of the adhoc nature of the SCMC approaches. In the present work, we use Ref.~\cite{jkb-mct}, which predicted dimension-dependent scaling exponents $\chi=(4-d)/6$ and $z=(8+d)/6$, where $\chi$ and $z$ are the roughness and dynamic exponents. To explore the rough phase in the model equations (\ref{kpz}) and (\ref{phieq}), we apply the SCMC framework developed in Ref.~\cite{jkb-mct}; see also Refs.~\cite{mhd-turbu,amit-ab-jkb,ab-jkb-mct,debayan2} in related contexts. With $\lambda=\lambda_1$, there are no vertex corrections in the model~\cite{abfrey1,ertas}, giving $\chi_h+z=2$,  an exact relation~\cite{sm-noise}, where we have assumed a single dynamical exponents for both $h$ and $\phi$, i.e., strong dynamic scaling; $\chi_h$ is the roughness exponent of $h$, $\chi_\phi$, the roughness exponent of $\phi$ remains unconstrained by the Galilean invariance. The SCMC calculations for the present problem starts with the scaling forms for the 
correlation and response functions for $h$ and $\phi$, which in the low frequency limit are 
\begin{eqnarray}
G_a^{-1}({\bf k},\omega)&=&-i\omega + \Gamma k^z,\\
 C_a({\bf k},\omega)&=&\langle |a({\bf k},\omega|^2\rangle = \frac{2D_a\Gamma k^{-2\chi_a-d-z}}{\omega^2 + \Gamma^2 k^{2z}},\\
 C_\times ({\bf k},\omega)&=&\langle h({\bf k},\omega)\phi(-{\bf k},-\omega)\rangle \nonumber \\&=&\frac{2iD_\times ({\bf k})k^{-\chi_h-\chi_\phi-d-z}}{\omega^2 + \Gamma^2 k^{2z}},
\end{eqnarray}
$a=h,\phi$. We have used the Lorentzian approximation for the correlation functions in the low frequency limit. We have assumed the same damping coefficient $\Gamma$, which is consistent with the assumed strong dynamic scaling and Prandtl number $P=1$ at the strong coupling fixed point as  speculated in Ref.~\cite{sm-noise}. The different one-loop diagrams are shown in Fig.~\ref{diagrams}. Our contention is that the dimensionless ratio $\Gamma^2/(D_h\lambda^2)$ may be calculated from the one-loop diagrammatic expansions of $G^{-1}({\bf k},\omega), C_h({\bf k},\omega), C_\phi({\bf k},\omega)$. We further assume that these diagrammatic expansions are dominated by the respective one-loop contributions. This can happen when $z<2$ and $\chi_h,\chi_\phi>0$.  We apply SCMC on both (\ref{kpz}) and (\ref{phieq}). The SCMC on (\ref{kpz}) is identical in form with that discussed in Ref.~\cite{jkb-mct}, which we revisit here. 

From the self-consistent equation for $\Sigma_h({\bf k},\omega)$ at $\omega=0$, we find
\begin{eqnarray}
 \Gamma k^z &=&\lambda^2 \int \frac{d^dq}{(2\pi)^d}\frac{d\Omega}{2pi}[({\bf k-q})\cdot {\bf q}] \nonumber \\ &\times &({\bf k\cdot q}) C_{hh}({\bf q},\Omega) G_{hh}({\bf k-q},\omega-\Omega).
\end{eqnarray}
Setting $\omega=0$ and shifting ${\bf q\rightarrow q+k}/2$, we get in the long wavelength limit $k\rightarrow 0$
\begin{equation}
 \Gamma k^z = k_d \frac{\lambda^2 D_h}{2\Gamma_h} k^{2-\chi_h}{d}.
\end{equation}
Then using $\chi_h+z=2$, we obtain 
\begin{equation}
 \frac{\Gamma^2}{\lambda^2 D_h}= \frac{k_d}{2d}.\label{gamma}
\end{equation}
Next, we use the self-consistent equation for the auto-correlation function of $h$. Evaluating $C_h({\bf k},\omega)=\langle |h({\bf k},\omega)|^2\rangle$ at $\omega=0$, we get
\begin{eqnarray}
 \frac{2D_h}{\Gamma_h}&=&\frac{2}{\Gamma_h^2 k^{2z}}\int \frac{d^dq}{(2\pi)^d}[{\bf q\cdot (k-q)}]^2 C_h({\bf q},\Omega) C_h ({\bf k-q},-\Omega)\nonumber \\&=& \frac{D_h^2}{2\Gamma_h^3k^{2z}}\int\frac{d^dq}{(2\pi)^d}q^{-3\chi_h -2d+2}, \label{ch}
\end{eqnarray}
where we have used $\chi_h+z=2$.

From the self-consistent equation for $C_\phi({\bf k},\omega=0)$, we obtain
\begin{eqnarray}
 \frac{\Gamma^2}{D_h\lambda^2}&=&\frac{1}{2}\frac{S_d}{(2\pi)^d}\frac{1}{d-2+2\chi_\phi+\chi_h}\left(1-\frac{D_\times^2}{D_h D_\phi}\right).\label{cphi}
\end{eqnarray}
Using (\ref{gamma}), (\ref{ch}) and (\ref{cphi}), we obtain
\begin{equation}
 2\chi_\phi = \left(1-\frac{D_\times^2}{D_h D_\phi}\right) d-d+2 -\chi_h,
\end{equation}
giving a relation between $\chi_h$ and $\chi_\phi$.  Now from (\ref{gamma}) and (\ref{ch}), we find~\cite{jkb-mct}  
\begin{equation}
 \chi_h=\frac{4-d}{6}.\label{chih}
\end{equation}
Using (\ref{chih}) then, we find
\begin{equation}
 \chi_\phi = \left(1-\frac{D_\times^2}{D_h D_\phi}\right)d - \frac{5d-8}{6},\label{chiphifinal}
\end{equation}
as the spatial scaling exponent of $\phi$. From (\ref{chiphifinal}), it is clear that $\chi_\phi$ reduces, as $|D_\times| $ rises, giving increasing suppression of $\phi$-fluctuations as the amplitude of the cross correlation function rises. Since the maximum value of $\mu\equiv D_\times^2/D_hD_phi$ is unity, we find the minimum of $\chi_\phi$ as
\begin{equation}
 \chi_{\phi}^\text{min}=- \frac{5d-8}{6}
\end{equation}
in $d$-dimensions. For instance, we find at $d=1$ and with $\mu=1/2$, $\chi_\phi=1$ and $\chi_{\phi}^\text{min}=1/2$; at $d=2$, $\mu =1/$, $\chi_\phi=2/3$ and $\chi_{\phi}^\text{min}=-1/3$.

We have thus investigated the effects of noise crosscorrelations on the strong coupling phases of the coupled model, where one field is autonomous and the second field is advected by it, in $d$-dimensions. By using a mode-coupling theory, we show that as the amplitude of the cross correlation function, which scales with the noise crosscorrelation amplitude, the spatial scaling or the roughness exponent of the second field monotonically decreases. Thus, the effect of noise crosscorrelation is to {\em reduce} the fluctuations in the second field.  In conjunction with the results reported in Ref.~\cite{sm-noise}, we can draw a coherent, broad picture. For instance, it was shown in Ref.~\cite{sm-noise} by using a dynamic RG framework that the second field at 1D, the second field is most rough when the noise crosscorrelation is entirely absent. This roughness decreases at a finite threshold for the noise crosscorrelation. It was speculated in Ref.~\cite{sm-noise} that in the putative strong coupling phase in 1D, inaccessible within a dynamic RG calculation, $\chi_\phi$ should be even less. This expectation is confirmed by the present work. In the higher dimensions, where there are no RG results available, our present results point towards similar trend for $\chi_\phi$ as a function of the cross correlation amplitude.  


\end{document}